\documentclass[superscriptaddress,noeprint,aps,reprint,floatfix]{revtex4-1}
\usepackage{graphicx}
\usepackage{graphics}
\usepackage{amssymb}
\usepackage{amsmath}
\usepackage{tikz}
\usepackage{dcolumn}
\usepackage[utf8]{inputenc}
\usepackage{bm}
\usepackage{color}
\usepackage{float}
\usepackage{hyperref}

\newcommand{\cn}[1]{\raisebox{.5pt}{\textcircled{\raisebox{-.9pt} {#1}}}}
\begin{document}
\title{Topology of many-body edge and extended quantum states in an open spin chain: 1/3--plateau, Kosterlitz-Thouless transition, and Luttinger liquid} 
\author{R. R. Montenegro-Filho}
\affiliation{Laborat\'{o}rio de F\'{i}sica Te\'{o}rica e Computacional,
Departamento de F\'{i}sica, Universidade Federal de Pernambuco, 50760-901 Recife-PE, Brasil}
\author{F. S. Matias}
\affiliation{Laborat\'{o}rio de F\'{i}sica Te\'{o}rica e Computacional,
Departamento de F\'{i}sica, Universidade Federal de Pernambuco, 50760-901 Recife-PE, Brasil}
\affiliation{Instituto de F\'{i}sica, Universidade Federal de Alagoas, 57072-970 Maceió-AL, Brasil}
\author{M. D. Coutinho-Filho}
\affiliation{Laborat\'{o}rio de F\'{i}sica Te\'{o}rica e Computacional,
Departamento de F\'{i}sica, Universidade Federal de Pernambuco, 50760-901 Recife-PE, Brasil}
\date{\today}

\begin{abstract}
Quantum many-body edge and extended magnon excitations from the  
1/3 -- plateau of the anisotropic Heisenberg model on an open AB$_2$ chain in a magnetic field $h$ 
are unveiled using the density matrix renormalization group and exact diagonalization. By tuning both the anisotropy and $h$ in the rich phase diagram, the edge states penetrate in the bulk, whose gap closes in a symmetry-protected 
topological Kosterlitz-Thouless transition. Also, we witness the squeezed chain effect, the breaking of the edge states degeneracy, and a topological change of the excitations from gapped magnons with quadratic long-wavelength dispersion to a linear spinon dispersion in the Luttinger liquid gapless phase as the anisotropy $\lambda$ approaches the critical point from the $\lambda>0$ side of the phase diagram. 
\end{abstract}

\maketitle
\section{Introduction}
Recently, increasing experimental and theoretical attention was given
to topological aspects of condensed matter physics \cite{Wen2019,*Sachdev2019}.
In one-dimensional (1D) systems, an early essential role of topology was provided by the 
so-called \textit{Haldane conjecture}\cite{nobelhaldane,PhysRevLett.50.1153,*Haldane1983}:
the ground state of integer (half-integer) spin chains is gapped (gapless). 
In fact, the conjecture was experimentally verified in spin-1 chains \cite{Buyers1986,*Tun1991};
further, density matrix renormalization group (DMRG) studies confirmed that
the \textit{bulk} gapped ground state displays spin-1/2 fractionalized \textit{edge} states 
in open chains \cite{White1992,*White1993}.
Topological insulators \cite{Hasan2010} share with these systems some general aspects \cite{Chen2011,Chen2013,Verresen2018}:
an insulating bulk and a conducting surface (edge states) are intrinsically connected, a phenomenon known as bulk-boundary correspondence. The Su-Schrieffer-Heeger (SSH) dimerized model \cite{Su1979}, and trimer models \cite{MartinezAlvarez2019}, including a diamond chain \cite{Pelegri2019}, are examples of models that
manifest the bulk-boundary correspondence in regions of their parameter space.
In addition, the phonon structures arising from mechanical isostatic \cite{Kane2014} and  Maxwell \cite{Mao2018}
lattices can be understood 
from the akin framework of topological band theory of electronic systems, including the bulk-boundary correspondence.
Also, chiral magnonic edge states in ferromagnetic skyrmion crystals controlled by
magnetic fields were reported \cite{PhysRevResearch.2.013231}.
Besides, we mention that the association of a two-dimensional Chern number with a one-dimensional system was also
suggested for photonic quasicrystals \cite{Kraus2012}, and fermionic systems in quasi-periodic optical
superlattices \cite{MartinezAlvarez2019,Lang2012}.

Gapped ground states of spin chains, either with spin-1 or more complex unit cells with spin-1/2 sites, imply  plateaus
in the magnetization ($m$) curves as a function of the magnetic field ($h$): $m(h)$. 
This is a topological quantization of the magnetization due to the 
presence of $h$, analogously to the quantum Hall effect \cite{OYAPrl97}. 
Recently, this issue was investigated in modulated spin 
chains \cite{Hu2014,*Hu2015}, with particular attention to the edge states of open systems. 
On the other hand, a magnetization plateau at 1/3 of the saturation magnetization (1/3 -- plateau) has been observed 
in several model systems.
The isotropic $AB_2$ chain exhibits a ferrimagnetic ground state 
\cite{Macedo1995,Tian1996,AlcarazandMa,PRL97Raposo,*PRB99Raposo} and the 1/3 -- plateau
in $m(h)$ \cite{PhysA2005,Coutinho-Filho2008}.  
The topological nature of the ground state manifests in topological Wess-Zumino terms of the non-linear sigma
model \cite{PRL97Raposo,*PRB99Raposo} or through its representation on a valence-bond
state basis \cite{Kolezhuk1997}. Likewise, the spin-(1/2,1) and spin-(1/2,5/2) alternating spin chains 
also exhibits a ferrimagnetic ground state, together with the 1/3 -- plateau \cite{AlcarazandMa,PhysRevB.55.8894,Maisinger1998,DaSilva2017}, and the 2/3 -- plateau \cite{Tenorio2011}, respectively. 
Besides, we mention the 1/3 -- plateau state of the quantum spin-1/2 XX diamond chain 
in a magnetic field \cite{Verkholyak2011}.
Further, in the
phase diagram of anisotropic spin models, the 1/3 -- plateau closes in a transition of the
Kosterlitz-Thouless (KT) type \cite{Kosterlitz1973,*Kosterlitz1974,*Kosterlitz2016,*nobelkosterlitz} as the
anisotropy changes \cite{YamamotoPRB99,Solid15Liu}. The KT transition is also observed in anisotropic ferrimagnetic
branched chains \cite{Verissimo2019,Karlova2019}. 
On the experimental side, the 1/3 -- plateau was observed in materials with three spin-1/2 sites 
per unit cell (diamond chain): the mineral azurite Cu$_3$(CO$_3$)$_2$(OH)$_2$ 
\cite{Kikuchi2005,Rule2008,Aimo2009,Rule2011,Jeschke2011}; and the compounds copper hydroxydiphosphate
Cu$_3$(P$_2$O$_6$OH)$_2$ \cite{Hase2006}, and alumoklyuchevskite K$_3$Cu$_3$AlO$_2$(SO$_4$)$_4$ 
\cite{Morita2017,*Fujihala2017}. Also, the 2/3 -- plateau was observed in 
a new mixed spin-(1/2,5/2) chain in a charge-transfer salt 
(4-Br-$o$-MePy-V)FeCl$_4$ \cite{Yamaguchi2020}.

In this work, DMRG and exact diagonalization (ED) results 
for open and closed anisotropic Heisenberg-$AB_2$ chains, respectively, unveil a very rich phase diagram and
related notable features. In particular, in open chains we identify a secondary plateau associated with
edge and extended magnon excitations from the 1/3--plateau. We stress that the edge magnon states that emerge 
from this plateau are many-body quantum states.
As one approaches the symmetry-protected [translational and $U(1)$ symmetries] topological quantum KT transition, the bulk penetration of
the edge states is enhanced, 
their degeneracy is broken, and the squeezed chain effect is observed. Further, at the KT transition and beyond, the
bulk magnon gap closes, while the edge states mix with the continuum and the Luttinger liquid (LL) excitations 
dominate the scenario.

In Sec. \ref{sec:pd}, we discuss the topology and phase diagram of the anisotropic Heisenberg-$AB_2$,
and a precise determination of the KT transition point. The edge states associated with the 1/3--plateau are
considered in Sec. \ref{sec:edge}, while 
gapped and gapless excitations around the topological KT transition are discussed in Sec. \ref{sec:bands}.
The boundary scattering length for the 1/3 -- plateau and the magnon-magnon scattering length for the 
fully polarized (FP) -- plateau magnons are reported in Sec. \ref{sec:sca}. A summary and conclusions are found in Sec. \ref{sec:summary}.  

\section{Topology and Phase diagram} 
\label{sec:pd}

\begin{figure}
\includegraphics*[width=0.47\textwidth]{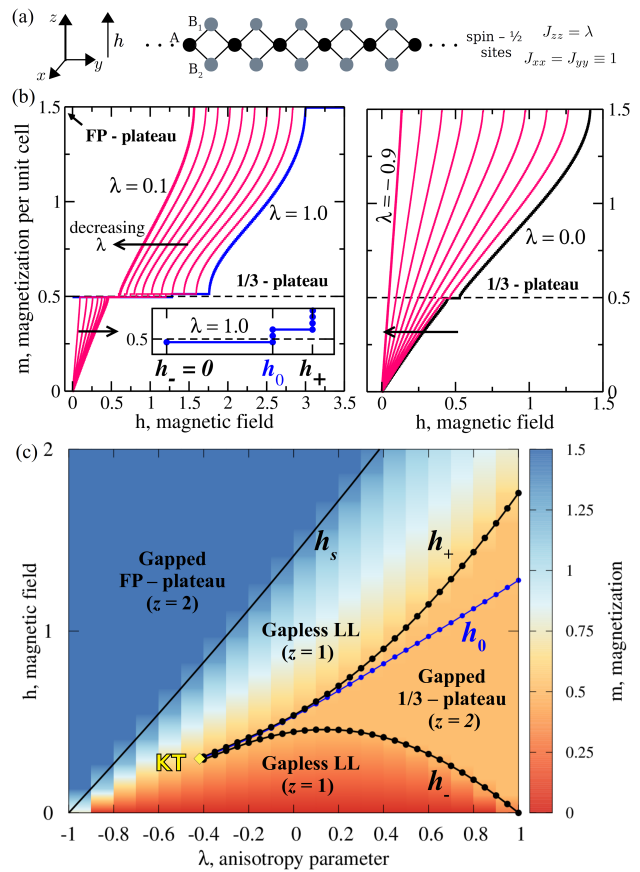}
\caption{(a) Schematic representation of the anisotropic Heisenberg Hamiltonian on the AB$_2$ spin-1/2 chain, 
under a magnetic field $h$. DMRG results for the open AB$_2$ chain with $N_c=121$ unit cells:  
(b) Magnetization per unit cell $m(h)$ for $1\geq\lambda\geq 0.1$ (left panel) and
$0.0\geq\lambda\geq -0.9$ (right panel), in 
steps of $\Delta \lambda=0.1$. Inset of the left panel: 
$m(h)$ for $\lambda=1.0$ in the vicinity of the 1/3 -- plateau bounded by $h_-=0$ and $h_{+}=1.76$, with 
a step at $h_0=1.28$; 
(c) Phase diagram: the color code refers to the $m$ values in (b). The exact critical line $h_s$ bounds
the FP -- plateau, while $h_-$, $h_0$, and $h_+$ are related to the 1/3 -- plateau. The gapped phases,
with dynamical exponent $z=2$, are separated by the gapless Luttinger liquid (LL) phase with $z=1$. The 1/3--plateau closes at
a Kosterlitz-Thouless (KT) transition: $\lambda_{KT}=-0.419\pm0.004$ and $h_{KT}=0.290\pm0.002$.}
\label{fig:mag}
\end{figure}

The anisotropic Heisenberg model on the $AB_2$ chain in an applied magnetic field $h$ reads:
\begin{eqnarray}
 H &=&\sum_{i=1}^{N_c}[S^x_{A,i}(S^x_{B,i}+S^x_{B,i-1})+S^y_{A,i}(S^y_{B,i}+S^y_{B,i-1})\nonumber\\
& &+\lambda S^z_{A,i}(S^z_{B,i}+S^z_{B,i-1})]-hS^z,
\label{eq:ham}
\end{eqnarray}
where $S^{x,y,z}_{B,i}=S^{x,y,z}_{B_1,i}+S^{x,y,z}_{B_2,i}$, $N_c$ is the number of unit cells of the system, the exchange couplings in the 
$xy$ plane define the unit of energy, $\lambda$  
is the exchange coupling in the $z$-direction, and 
$S^z=\sum_{i=1}^{N_c}(S^z_{A,i}+S^z_{B_1,i}+S^z_{B_2,i})$ 
is the  $z$ component of the total spin of the system, as illustrated in Fig. \ref{fig:mag}(a). 
We use DMRG to study open chains of $N_c$ unit cells,
with one $A$ site at each boundary, retaining 243 states per block and performing 12 sweeps in each calculation, such that the higher discarded weight was of order $10^{-9}$. We also study closed systems with $N_c=10$ and $N_c=12$ through ED.
The magnetization curves are obtained from the lowest energy in each total spin 
$S^z$ sector and $h=0$: $E(S^z)$,
since the Zeeman term in the Hamiltonian (\ref{eq:ham}) implies $E_{h}(S^z)=E(S^z)-hS^z$
for $h\neq0$. In a finite size system, the $m(h)$ curve is composed of finite size 
steps of width $\Delta h(S^z)$ at total spin $S^z$. Considering 
$h_{S^z+}$ and $h_{S^z-}$ as the extreme points of these steps, such that 
$\Delta h(S^z)=h_{S^z+}-h_{S^z-}$, we thus have $h_{S^z\pm}=\pm[E(S^z\pm1)-E(S^z)]$.
If $S^z$ is not at a thermodynamic-limit magnetization plateau state, we have 
$\Delta h(S^z)\rightarrow 0$ as $N_c\rightarrow\infty$, otherwise $\Delta h(S^z)\neq 0$ 
as $N_c\rightarrow\infty$. 

In Fig. \ref{fig:mag}(b) we present DMRG results ($N_c=121$) for $m(h)$ and the 
anisotropy in the interval $-0.9\leq \lambda \leq1$. 
The $m(h)$ curves display the FP -- plateau at the thermodynamic-limit (bulk) saturation
magnetization $m_s=3/2$, a plateau slightly below the bulk 1/3 -- plateau at $m_s/3=1/2$, and 
a secondary plateau, as shown in the inset for $\lambda=1.0$. The fields $h_-$, $h_0$ and $h_+$ define the 
width of the plateaus: the secondary one is associated with edge and extended magnon excitations from the 1/3--plateau. Here, these excitations will be examined in detail around the KT transition, in which case LL excitations also take place. In fact, in Fig. \ref{fig:mag}(c), a rich 
$h$-$\lambda$ phase diagram exhibits the various phases that play a significant role in our analysis.

In bulk, without broken translational symmetry, the possible occurrence of a plateau in $m(h)$ must satisfy 
the topological criterion \cite{OYAPrl97}: 
\begin{equation}
S_{c}-m=\text{integer}, 
\end{equation}
where $S_c$ is the maximum spin of a unit cell.
In our model, $S_{c}=3/2$, $m=1/2$ for the 1/3 -- plateau and $m=3/2$ for the FP -- plateau. 
Also, this topological criterion can be 
related \cite{Hu2014,Hu2015} to a Chern number $C_m$ defined in the two-dimensional parameter space
of an associated periodically modulated closed system under a twisted boundary condition.
Indeed, an $m$-plateau obeys the relation: 
\begin{equation}
C_m=-(S_c-m), 
\end{equation}
for $m\geq0$, with $C_{m}=-C_{-m}$ for $m<0$, i. e., $h<0$ not shown in Fig. \ref{fig:mag}. Thus, the FP -- plateau has a Chern
number $C_{3/2}=0$ and is a trivial insulating state; 
while the 1/3--plateau is a topological insulator with $C_{1/2}=-1$. 
In Sec. \ref{secsec:fpsca}, we present a detailed discussion of the trivial
insulating FP -- plateau state.  

In our open finite-size chain, a remarkable feature is the presence of edge states, leading to the splitting of the 1/3 -- plateau
into two plateaus. Consider, for example, the isotropic case shown in the inset of Fig. \ref{fig:mag}(b). The bulk
1/3 -- plateau has extreme points 
at $h_{-}=0$ and $h_{+}=1.76$ (for both spin-(1/2,1) \cite{PhysRevB.57.13610} and AB$_2$ \cite{PhysA2005} chains). However, in
the open finite-size system and $h_0\leq h<h_{+}$, the
magnon excitations occupy edge states inside the gap between the lower and upper bulk
band states and give rise to the two plateaus in $m(h)$. The transition between these
two plateaus occurs at $h_0=1.28$ for $\lambda=1$. 

The phase diagram of the AB$_2$-chain with $N_c=121$ unit cells is shown in Fig. \ref{fig:mag}(c). 
The extreme lines of the bulk plateaus, $h_-(\lambda)$, $h_+(\lambda)$, and $h_s(\lambda)$, are quantum critical
lines separating
a gapped insulating phase from the gapless LL phase, with dynamic critical exponent $z=2$ and $z=1$, respectively. 
The FP -- plateau
is bounded by $h_s(\lambda)=\frac{3\lambda}{2}+\frac{1}{2}\sqrt{8+\lambda^2}$, 
since the energy of the exact Goldstone mode (a $\Delta S^z=-1$ magnon) associated with this line reads:  
$\varepsilon_{\text{FP}}(k)=-\frac{3\lambda}{2}-\frac{1}{2}\sqrt{\lambda^2+8\cos^2(k/2)}+h$.
Therefore, for $h$ close to $h_s(\lambda)$, a high-dilute regime of magnons is verified, 
with the following low-lying excitation energy:
\begin{equation}
 \varepsilon(k)=-\mu+\frac{v^2 k^2}{2h_s},
 \label{eq:diluteregime}
\end{equation}
where $\mu=h_s-h$ and the spin-wave velocity is  
\begin{equation}
v=\frac{1}{\sqrt{2\left(1-\frac{3\lambda}{2h_{s}}\right)}}.
\label{eq:vfp}
\end{equation}
  
In addition, the 1/3 -- plateau is bounded by the critical lines $h_{-}(\lambda)$ and $h_{+}(\lambda)$, with a width 
$\Delta(\lambda)=h_+(\lambda)-h_-(\lambda)$. The plateau width $\Delta (\lambda)$ is the \textit{bulk gap} that
separates the two regions of the gapless LL phase: one with $m<1/2$, and the other with $m>1/2$, for the
same value of $\lambda$.
On the other hand, the low-energy theory of magnons in a gapped system under a magnetic field is
that of a Lieb-Liniger \cite{Lieb1963} Bose fluid with $\delta$-function interactions \cite{Affleck91}. In addition, in the
high dilute
regime of magnons, the theory is equivalent to a Tonks-Girardeau \cite{Tonks,*Girardeau} Bose system with a hard-core
repulsion \cite{Affleck91} or a fermionic system \cite{Tsvelik90,Affleck91,Montenegro-Filho2008,Tenorio2011}. 
Thereby in the high-dilute regime $h\rightarrow h_-\text{ or }h_+$, the low-energy magnon excitations
from the 1/3--plateau have dispersion relations as in Eq. (\ref{eq:diluteregime}), with $\mu=\pm (h-h_{\pm})$. 
For $h\lesssim h_{-}$, the magnons carry spin $\Delta S^z=-1$, while
for $h\gtrsim h_{+}$, the excitations carry spin $\Delta S^z=+1$. 
The $\Delta S^z=-1$ excitations can thus be understood as holes, in the 
reciprocal $q$-space, in a filled band of $\Delta S^z =+1$ hard-core magnons, and the bulk gap $\Delta(\lambda)$ 
is the particle-hole gap. The plateau closes at the KT quantum critical point: 
$\lambda_{KT}=-0.419\pm0.004$ and $h_{KT}=0.290\pm0.002$, estimated through the procedure described below.

\subsection{Kosterlitz-Thouless transition point: $\lambda_{KT}$ and $h_{KT}$}

In the LL gapless phase shown in Fig. \ref{fig:mag}(c), the transverse spin correlation function
should obey the asymptotic power-law behavior given by \cite{giamarchi2003quantum}
\begin{equation}
\Gamma(r)\sim \frac{1}{r^\frac{1}{2K}},
 \label{eqG}
\end{equation}
where $r$ is 
the distance between spins and $K$ is the Luttinger liquid parameter $K$, which 
depends on $h$ (or $m$) and $\lambda$. 
In the Kosterlitz-Thouless transition, the magnetization has 
the fixed value $m=1/2$ and the transition is induced by changing $\lambda$. 
In this case, $K=2$ at the critical point $\lambda=\lambda_{KT}$. 

We estimate the value of $\lambda_{KT}$ through a method successfully used to estimate the KT
transition points in a one-dimensional Bose-Hubbard model in Ref. \cite{Kuhner}.
In our case, the procedure consists in identifying the values of $\lambda$ 
at which $K=2$ for $m=1/2$ in finite size systems, and extrapolating the results to $N_c\rightarrow\infty$. 
We calculate the transverse spin correlation functions as
\begin{equation}
 \Gamma(r)\equiv\langle\langle S^+(l)S^-(l+r)\rangle\rangle_l, 
\end{equation}
where the $\langle\langle \ldots \rangle\rangle_l$ indicate the quantum expectation value 
and an average of the correlation over all pairs 
of cells with a distance $l$ between then, in order to minimize the effects of the
open boundaries of the chain. 
\begin{figure}
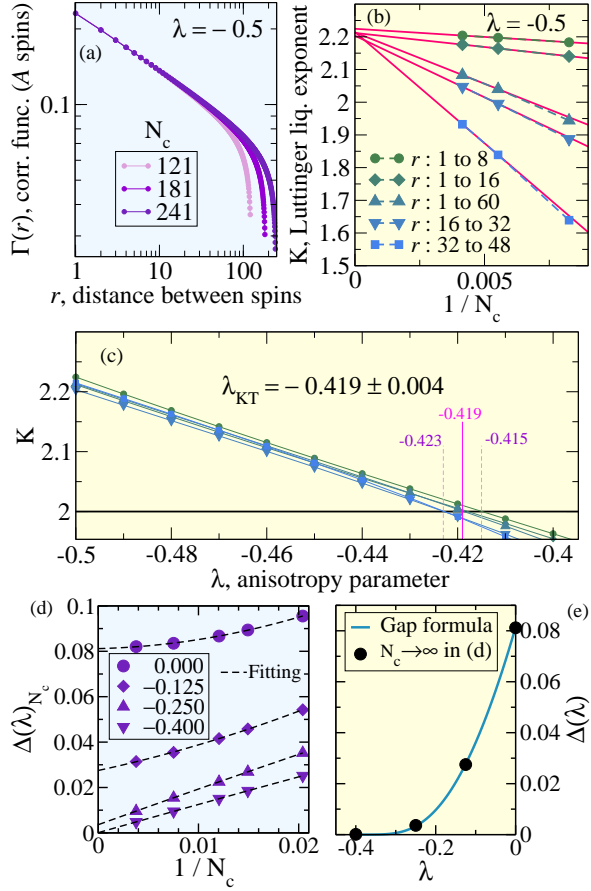

 \includegraphics*[width=0.43\textwidth]{fig2.eps}
 \includegraphics*[width=0.43\textwidth]{fig2de.eps}
 \caption{Critical $\lambda$ of the Kosterlitz-Thouless transition: $\lambda_{KT}$. 
 (a) Transverse spin correlation functions $\Gamma(r)=\langle\langle S^+(l)S^-(l+r)\rangle\rangle_l$ 
 between $A$ spins as a function of distance $r$ for $\lambda=-0.5$ at the magnetization ($m$) of
 the 1/3--plateau: $m=(1/2)-(1/2N_c)$, for the number of unit cells indicated. For a given system size,
 $\Gamma(r)$ is calculated by averaging over all pairs of spins separated by the distance $r$. (b) Luttinger liquid
 exponent $K$ as a function of $1/N_c$ 
 for the three system sizes shown in (a) and $\lambda=-0.5$.  The value of $K$ is determined by fitting $\Gamma(r)$
 to the expected long-distance power-law behavior $1/r^{1/2K}$ through the indicated intervals of $r$. 
 Full lines are linear extrapolations of $K$ to $N_c\rightarrow\infty$, by considering the two highest system sizes.
 (c) Extrapolated value of $K$ as a function of $\lambda$ for each fitting interval 
 indicated in (b). The critical $\lambda$ is estimated from the minimum and maximum values of 
 $\lambda$ at which $K=2$, within the set of investigated fitting intervals.
 (d) 1/3 -- plateau width $\Delta(\lambda)_{N_c}$ as a function of $1/N_c$ for 
the indicated values of $\lambda$, dashed lines are fittings to a polynomial expression. 
(e) ($\bullet$) $\Delta(\lambda)$ from (d) as a function of $\lambda$. The full line is the
fitting of this data to the essential singularity formula $A\exp{\left(B/\sqrt{\lambda-\lambda_{KT}}\right)}$.}
 \label{fig:lc}
\end{figure}

In Fig. \ref{fig:lc}(a), we show $\Gamma(r)$ between $A$ spins for $\lambda=-0.5$ and $N_c=121,181,\text{ and }241$, 
at $m=1/2-(1/2N_c)$.
For each system size, we fit the data in different intervals of $r$ to the asymptotic expression in Eq. (\ref{eqG}). The following intervals were considered for $r$: $[1,8]$; $[1,16]$; $[1,60]$; $[16,32]$; 
and $[32,48]$ for values of $\lambda$ around the KT transition. In particular, in Fig. \ref{fig:lc}(b) we show  $K$ as
a function of the system size for $\lambda=-0.5$ and the chosen $r$-intervals. We see that a straight line can be a good scale 
function for $K$ in all studied $r$-intervals. Hence, we fit a linear function to the 
data of the two largest system sizes in order to obtain very confident extrapolated value of $K$, i. e.,
with very little dispersion. Indeed, for the case shown in Fig. \ref{fig:lc}(b), $\lambda=-0.5$, the extrapolated
value of $K$ is in the range $2.218 \pm 0.006$. In Fig. \ref{fig:lc}(c), we show the extrapolated values of $K$ as
a function of $\lambda$ for each of the chosen $r$-intervals. The KT critical value of $\lambda$: 
\begin{equation}
\lambda_{KT}=-0.419\pm0.004, 
\end{equation}
 is estimated by considering the minimum and maximum values of $\lambda$ at which $K=2$, in all chosen $r$-intervals. 
The bulk gap $\Delta (\lambda)$ nullifies following an essential singularity form
\begin{equation}
 \Delta(\lambda)=A\exp{\frac{B}{\sqrt{\lambda-\lambda_{KT}}}},\label{essenSin}
\end{equation}
where $A$ and $B$ are constants. In Fig. \ref{fig:lc}(d) we show a scale analysis of the plateau width 
for some values of $\lambda$ in the gapped phase. In Fig. \ref{fig:lc}(e), we present the extrapolated 
values of the bulk gap as a function of $\lambda$ and the fitting of them to the expression (\ref{essenSin}).
 
 \begin{figure}
 \includegraphics*[width=0.40\textwidth]{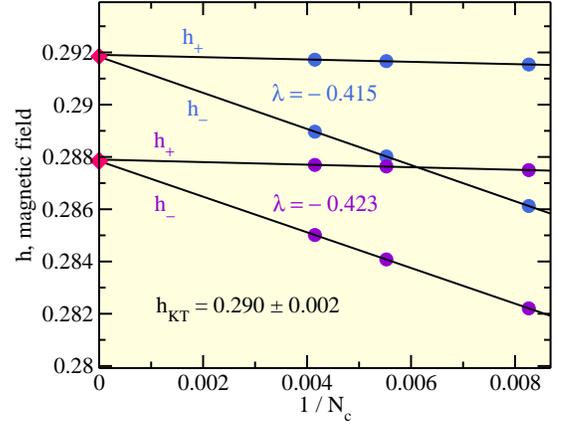}
 \caption{Critical $h$ of the KT transition: $h_{KT}$. 
 Extreme fields of the finite-size 1/3 -- plateau magnetization: $m=(1/2)-(1/2N_c)$, as 
 a function of $1/N_c$ for $\lambda=-0.415$ and $\lambda=-0.423$, which are the estimated minimum and 
 maximum values of $\lambda$ at the KT transition. 
For each value of $\lambda$, we use a linear extrapolation in $1/N_c$ to evaluate
the thermodynamic-value of $h$ for $m=1/2$. The critical field is estimated
as the average of the extrapolated values.}
 \label{fig:hc}
\end{figure}

The value of the critical field $h_{KT}$ can be estimated by a scaling analysis of the extreme fields $h_{-}$ and $h_{+}$ of the 
finite-size 1/3 -- plateau magnetization at $m=1/2-(1/2N_c)$. In Fig. \ref{fig:hc}, we 
present $h_{-}$ and $h_{+}$ as a function of system size for the minimum and maximum values of $\lambda_{KT}$: -0.415 and -0.423.
In both cases, an excellent linear scale function fits the data for $h_-$ and $h_+$.
For $\lambda=-0.415$, the extrapolated values of $h_{-}$ and $h_{+}$ 
differ by $7\times10^{-5}$; while for $\lambda=-0.419$, the difference is $5\times10^{-5}$. We estimate the critical 
field of the KT transition, $h_{KT}$, as the range from the extrapolated value of $h_{-}$ at $\lambda=-0.423$ to
the extrapolated value of $h_{+}$ at $\lambda=-0.415$, thus obtaining: 
\begin{equation}
h_{KT}=0.290\pm0.002. 
\end{equation}

The $AB_2$ anisotropic chain is invariant under the exchange of the two $B$ sites of a unit cell, so 
the Hamiltonian does not connect the singlet and triplet states of these pairs. The localized singlet pairs appear
in higher energy states of the system that are not activated by either the magnetic field nor the anisotropy.
Thus, the $h\text{ vs. }\lambda$ phase diagram of the $AB_2$ anisotropic chain is the same as that   
of the alternating spin-(1/2,1) anisotropic chain \cite{YamamotoPRB99,Solid15Liu}, 
and we can compare the results for this chain with our estimates for $\lambda_{KT}=-0.419\pm0.004$ and $h_{KT}=0.290\pm0.002$. 
These values disagree with the ones suggested for the anisotropic alternating chain 
in Ref. \cite{Solid15Liu} by observing the behavior
of the two-site entanglement calculated by the infinite time-evolving block-decimation (iTEBD)
algorithm: $\lambda=-0.53$ and $h=0.23$. On the other hand, the values estimated in Ref. \cite{YamamotoPRB99} through a finite size analysis of the central charge and plateau size: $\lambda=-0.41\pm0.01$ and $h=0.293$, are compatible with our more precise results.  
 
\section{Edge magnon excitations of the gapped 1/3 -- plateau}
\label{sec:edge}

\begin{figure}
\includegraphics*[width=0.47\textwidth]{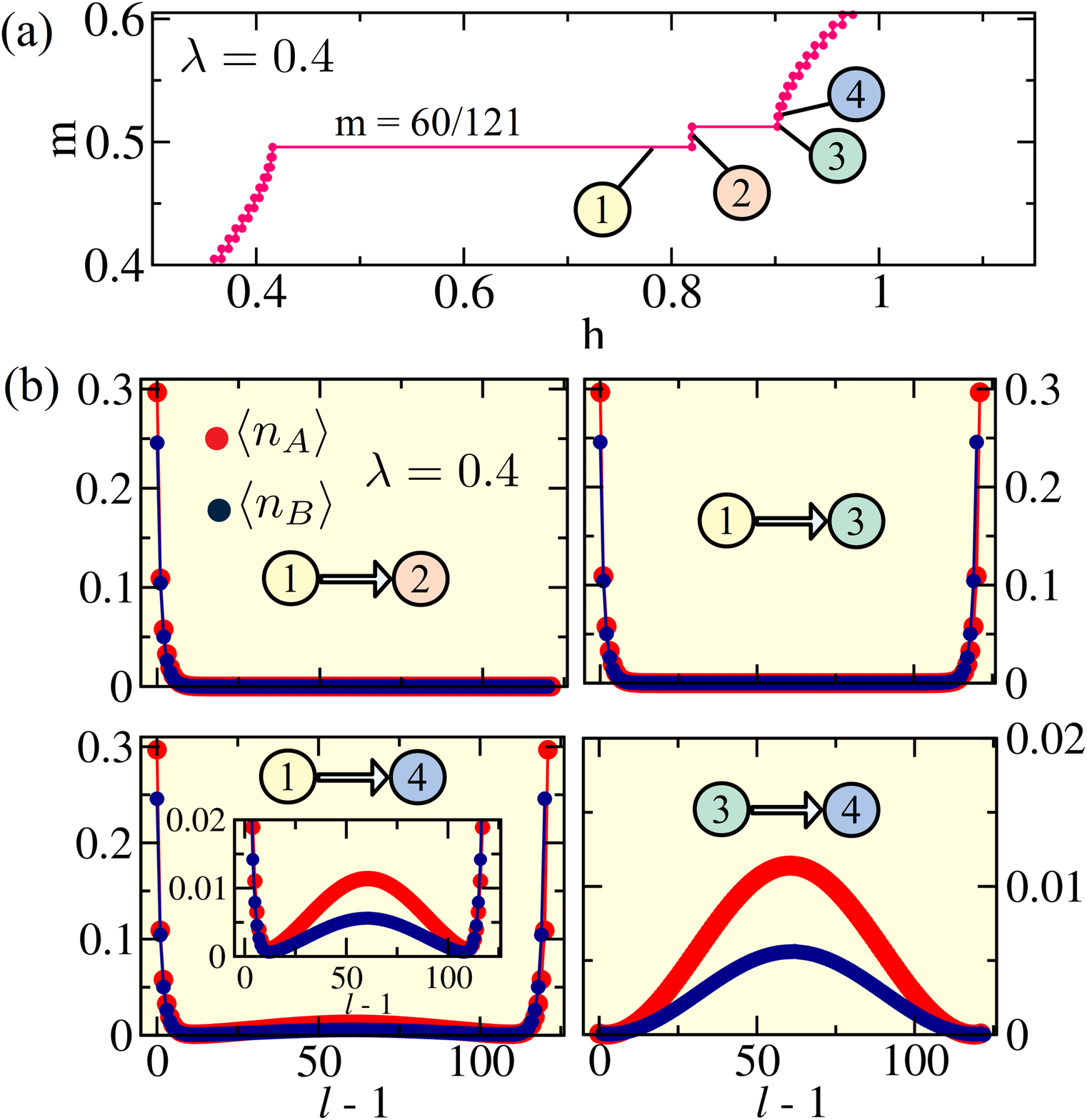}
\caption{DMRG results for $m(h)$ and the average magnon distribution along the AB$_2$ open chain with $N_c=121$, at $\lambda=0.4$. 
(a) $m(h)$ in the vicinity of the 1/3 -- plateau displaying the indicated $m$-states: \cn{1} ($m=60/121$),
\cn{2} ($m=61/121$), and \cn{3} ($m=62/121$); and the first gapless $m$-state above the plateau (onset of the continuum): 
\cn{4} ($m=63/121$).
(b) Average magnon distribution at sites $A$, $\langle n_A\rangle$, and $B$,
$\langle n_B\rangle\equiv\langle n_{B_1}\rangle+\langle n_{B_2}\rangle$, as a function of cell
position $l-1$. Excitations $\cn{1}\rightarrow\cn{2}$,
$\cn{1}\rightarrow\cn{3}$, and $\cn{1}\rightarrow\cn{4}$ create 1, 2, and 3 magnons above the $m$-state \cn{1};
while $\cn{3}\rightarrow\cn{4}$ creates one magnon in the $m$-state \cn{3}.}
\label{fig:edgea}
\end{figure}

In our open chain, the topological quantum
phase transition
from the insulating ($z=2$) to the metallic phase ($z=1$) manifests in the penetration into the bulk of the
edge (surface) states \cite{Griffith2018,*Rufo2019}. We start by discussing
the magnon edge states associated with the topological insulator at the 1/3 -- plateau in 
the open AB$_2$-chain of size $N_c=121$ and $\lambda=0.4$. In Fig. \ref{fig:edgea}(a) we present $m(h)$ in the vicinity of the 1/3 -- plateau ($m=0.5$ in the thermodynamic limit). 
In this finite-size system, the $m$-states that characterize the 1/3 -- plateau phase are
labeled by \cn{1} ($m=60/121$), \cn{2} ($m=61/121$), and \cn{3} ($m=62/121$); while the first extended state
above the plateau is labeled by \cn{4} ($m=63/121$). 
As $m$ changes from a state $\cn{i}$ to a state $\cn{f}$, the change in the average distribution of $\Delta S^z=+1$ magnons on
sites $A$, $\langle n_A\rangle$, and sites $B=B_1+B_2$, $\langle n_B\rangle$, are calculated through 
$\langle n_X\rangle_{\cn{i}\rightarrow\cn{f}}=\langle S^z_X\rangle_{f}-\langle S^z_X\rangle_{i}$,
with $X=A\text{ or }B$, as shown in the panels of Fig. \ref{fig:edgea}(b).  
In panel $\cn{1}\rightarrow\cn{2}$, the magnon distribution indicates that a magnon added to the
state $\cn{1}$ is localized at the left edge of the chain; while 
a second magnon added to $\cn{1}$, panel $\cn{1}\rightarrow\cn{3}$, is localized at the right edge. Thus, the 
distributions of one- and two-magnon states above $\cn{1}$ indicate the presence of localized states at
both edges of the chain, implied by the inversion
symmetry of the finite-size chain relative to its center, with the density on $A$ sites higher than those on $B$ sites. 
Concerning the three-magnon state, panel $\cn{1}\rightarrow\cn{4}$ in Fig. \ref{fig:edgea}(b), the magnon 
distribution evidences that the third magnon occupies a metallic state, which extends throughout the bulk. 
Indeed, panel $\cn{3}\rightarrow\cn{4}$ in Fig. \ref{fig:edgea}(b) presents the distribution of
this one-magnon extended state, which is clearly isolated from the edge states. In Appendix \ref{sec:appendixA}
we show that the magnetization and magnon distributions for an even number of unit cells and the same 
boundary conditions have the same physical features; while using a boundary condition with a $B_1,B_2$ 
at one extreme gives rise to only one edge state. Further, in Appendix \ref{sec:appendixB} we present the 
average local magnetizations along the chain, from which the magnon distributions were calculated.

\begin{figure}
\includegraphics*[width=0.47\textwidth]{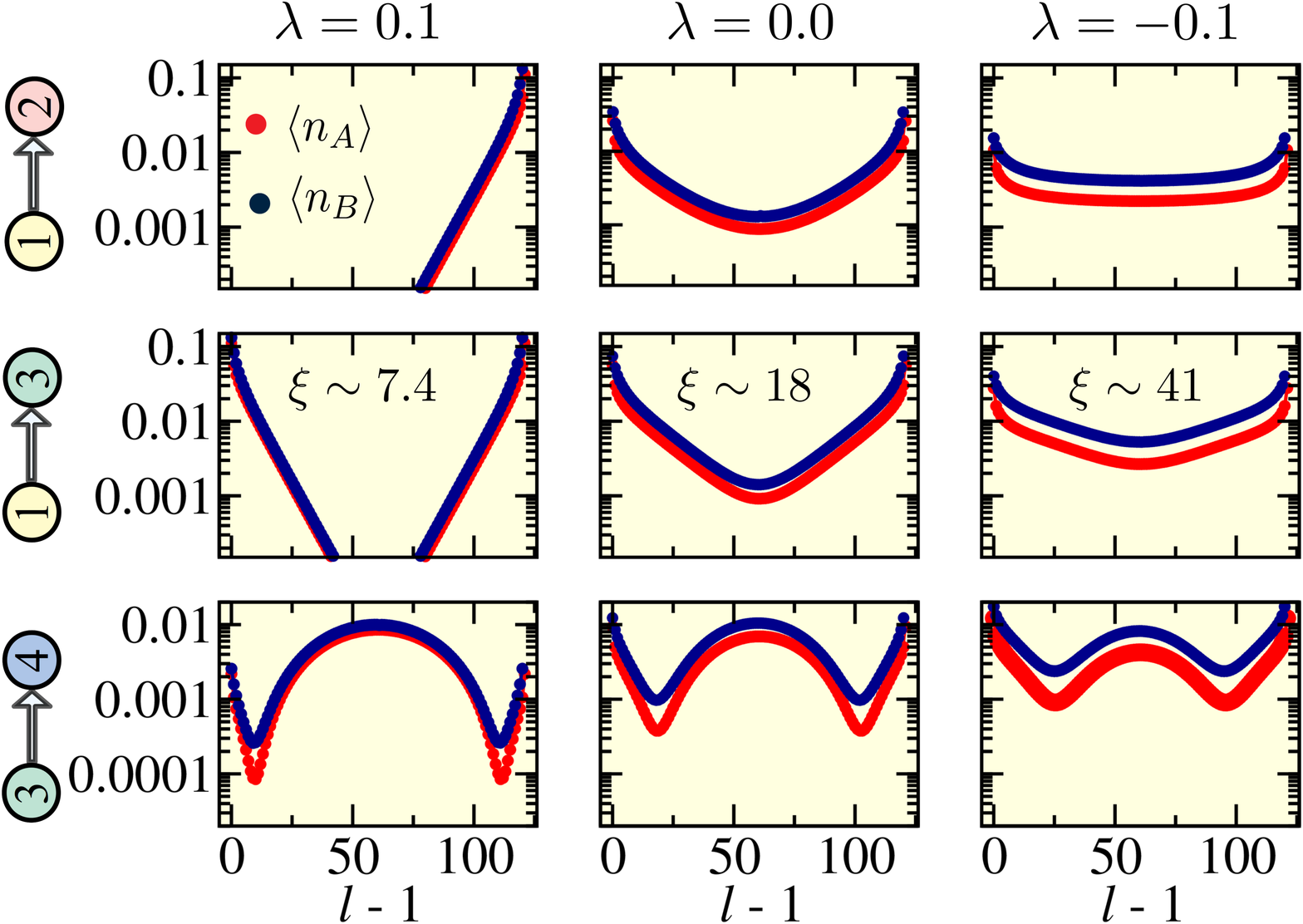}
\caption{DMRG results for the average magnon distributions $\langle n_A\rangle$ and $\langle n_B\rangle$ along the AB$_2$ open chain
with $N_c=121$, as the KT transition gets closer. A log-normal scale is used in the figures. The panel columns are data for
$\lambda=-0.1,~0.0,~\text{ and }0.1$, from left to right; while
panel lines show $\langle n_A\rangle$ and $\langle n_B\rangle$ for the $\cn{1}\rightarrow\cn{2}$, $\cn{1}\rightarrow\cn{3}$, and
$\cn{3}\rightarrow\cn{4}$ excitations. The localization length $\xi$ shown 
in the second line of the panels is obtained by fitting the data of $\langle n_A\rangle$ in the range
$30\leq x \leq 40$ to $\text{e}^{-x/\xi}$, with cell position $x=l-1$.}
\label{fig:edgeb}
\end{figure}

Now, we shall focus on the very interesting behavior of edge and bulk magnon excitations as the 1/3 -- plateau gets 
closer to the KT critical point: $\lambda_{KT},h_{KT}$. In Fig. \ref{fig:edgeb} (semi-log plots), we 
present the average distributions of one ($\cn{1}\rightarrow\cn{2}$) and two ($\cn{1}\rightarrow\cn{3}$) magnon
excitations above $\cn{1}$, as well as the isolated one-magnon extended state (excitation $\cn{3}\rightarrow\cn{4}$),
for $\lambda=0.1$, $0.0$, and $-0.1$, corresponding to the first, second, and third columns, respectively. For $\lambda=0.1$
(first column) the one-magnon state is exponentially localized at the right edge, while the two-magnon state displays
one localized magnon at each edge, similarly to the $\lambda=0.4$ case in Fig. \ref{fig:edgea}(b). Thus, left and right
edge states are still degenerate. However, at $\lambda=0$ (second column), the gap between the two edge states
[$\equiv \Delta h=6\times 10^{-4}$, as shown in Fig. \ref{fig:edgebb}(a)] is open and the one-magnon state
displays a symmetrical density on \textit{both edges} of the chain due to hybridization, thus leading to
bulk penetration. Also, the two-magnon state exhibits similar behavior with a small dip
at the center of the chain. Further, as shown in Fig. \ref{fig:edgeb}, as the bulk gap $\Delta(\lambda)$ 
(width of the 1/3 -- plateau) decreases the localization length $\xi$ of
the edge states 
increases, since $\xi(\lambda)\sim 1/\Delta(\lambda)$, and the edge state becomes more extended. In fact, 
for $\lambda=-0.1$, 
the density profile of one- and two-magnon edge states are very extended, 
with the density at the boundaries approaching their values in bulk. Using data 
from the excitation $\cn{1}\rightarrow\cn{3}$ in Fig. \ref{fig:edgeb} for $\lambda=0.1$, $0.0$, and $-0.1$, 
we have estimated the values of the localization length: $\xi=7.4,~18,\text{ and }41$, respectively. 
On the other hand, for $\lambda=0.1$, the weight at the boundaries of the isolated one-magnon extended state, excitation $\cn{3}\rightarrow\cn{4}$,  
is much higher than the practically negligible weight in the 
$\lambda=0.4$ case [see Fig. \ref{fig:edgea}(b)]. 
In fact, as the gap closes, the insulating bulk is squeezed, as
shown in Fig. \ref{fig:edgeb} by the decreasing of the distance between the two minima in the $\cn{3}\rightarrow\cn{4}$
excitation, and also by the increasing penetration of the edge states for the two-magnon
$\cn{1}\rightarrow\cn{3}$ state.  
Notably, far enough from the boundaries, the bulk wavefunction
of the $\cn{3}\rightarrow\cn{4}$ one-magnon state is that of a squeezed chain of size $L-2a_b$, where $a_b$ is
the boundary scattering length of an effective
repulsive potential \cite{Sca2}. A more detailed quantitative discussion 
is presented in Sec. \ref{secsec:bsca}.

\section{Gapped and gapless excitations around the topological KT transition}
\label{sec:bands}

\begin{figure}
\includegraphics*[width=0.47\textwidth]{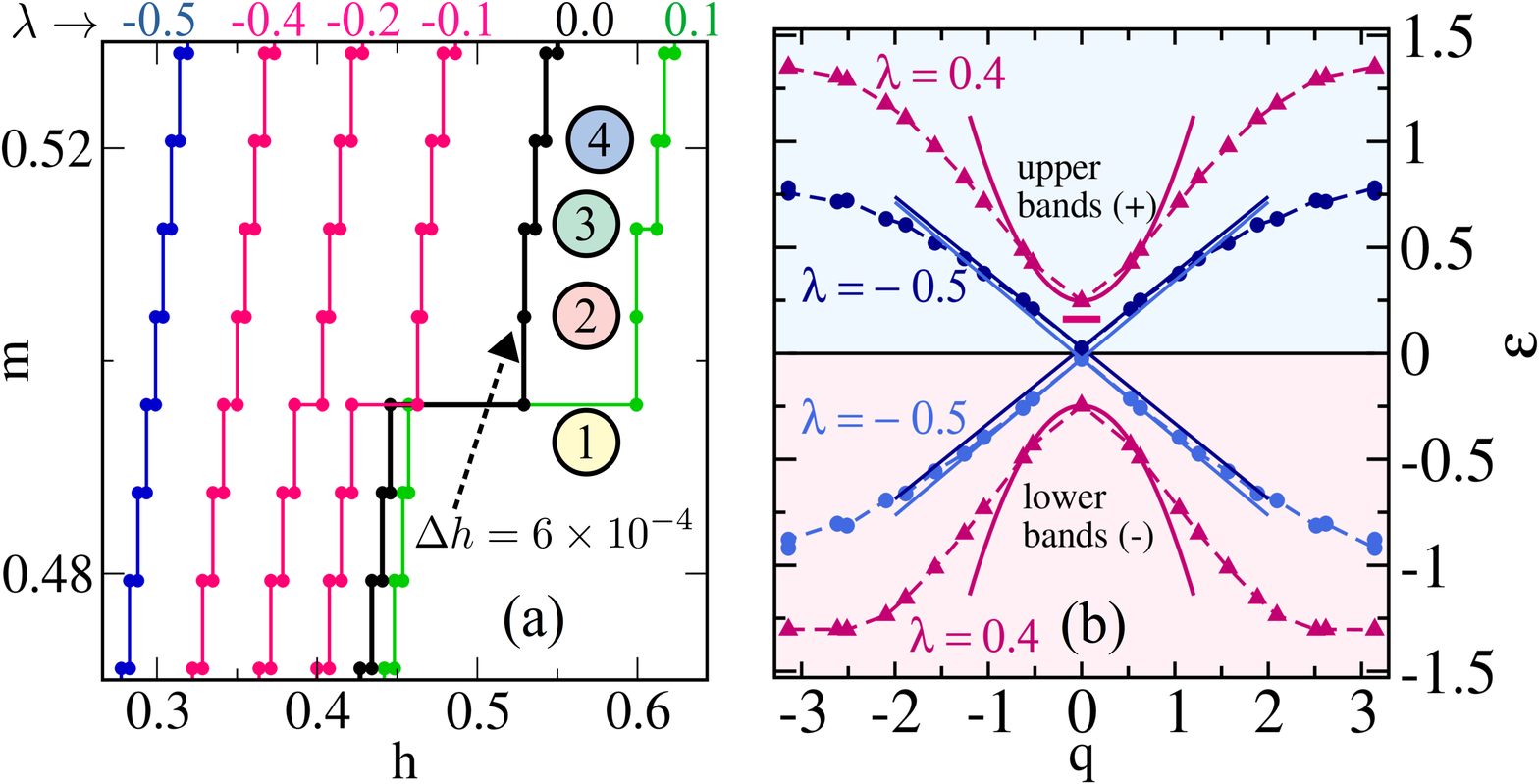}
\caption{(a) DMRG results for $m(h)$ in the vicinity of the 1/3--plateau of the AB$_2$ open
chain with $N_c=121$ for the indicated values of $\lambda$
and the indicated $m$-states: \cn{1} ($m=60/121$),
\cn{2} ($m=61/121$), \cn{3} ($m=62/121$), and \cn{4} ($m=63/121$), as in Fig. \ref{fig:edgea}. Notably, 
for $\lambda=0.0$, there is a finite-size step of size $\Delta h = 6\times 10^{-4}$ at the $m$-state \cn{2}.
(b) Upper and lower band energies for 
$\Delta S^z =+1$ magnons of wave-vector $q$, with $h$ at the center of the
1/3 -- plateau, $(h_{+}+h_{-})/2$, for $\lambda=0.4$ (\textcolor{red}{$\blacktriangle$}) and
$-0.5$ (\textcolor{blue}{$\bullet$}), using ED results from $N_c=10$ and $N_c=12$ under 
closed boundary conditions. 
We also indicate the two-fold degenerate magnon edge states (\textcolor{red}{\textbf{--}}) below the bottom 
of the magnon upper band, using DMRG for $\lambda=0.4$ and an open chain with $N_c=121$. 
}
\label{fig:edgebb}
\end{figure}

In Fig. \ref{fig:edgebb}(a), we show $m(h)$ in the vicinity of the 1/3 -- plateau
for the indicated values of $\lambda$ and using the same state labeling of Figs. \ref{fig:edgea} and \ref{fig:edgeb}. 
A remarkable feature is the \textit{breaking of the degeneracy} between states 
$\cn{2}$ and $\cn{3}$ for $\lambda\sim 0.0$ (black curve), as one decreases $\lambda$ from
$\lambda=0.1$ (green curve), in accord with the magnon distribution in Fig. \ref{fig:edgeb}. 
In fact, for $\lambda=0.0$, there is a gap of size $6\times10^{-4}$ between these states, 
implying a $m$-step of width $\Delta h =6\times10^{-4}$ in the $m$-state $\cn{2}$. 
Further, the width of the $m$-step increases (decreases) at the $m$-state $\cn{2}$ ($\cn{1}$) as
the gap closes and all states take part of the continuum at the KT critical point $(\lambda_{KT},h_{KT})$ in
the thermodynamic limit. 
Accordingly, in our finite-size system we observe uniformity in the values of the widths of the
$m$-steps, as shown in Fig. \ref{fig:edgebb}(a) for $\lambda=-0.5$ (blue curve), 
a signature of a gapless LL phase. 

In addition, a remarkable topological change in the dispersion relation of the low-energy magnetic
excitations takes place around the KT critical point.
There are two kinds of bulk magnetic excitations from the 1/3 -- plateau: one carrying a spin $\Delta S^z=+1$, which
increases the 1/3 -- plateau total spin $S^{z}_{1/3}$ by one unit; and the other, carrying a spin $\Delta S^z =-1$,
which decreases $S^{z}_{1/3}$ by one unit. The excitations with $\Delta S^z=-1$ can be understood as a hole, in the 
reciprocal $q$-space, in a filled band of $\Delta S^z =+1$ excitations. The magnetic field acts as a chemical
potential: for $h=h_{-}$ the lower band
is filled and the upper one is empty; increasing $h$, the magnetization does not change (plateau region) up to
$h=h_+$, at which the upper band starts to be filled. Defining $E_{1/3}$ as the total energy of the 1/3 -- plateau
magnetization and $h=0$, the energy $\varepsilon_{\pm}(q)$ of the upper (+) and lower (-) bands are given
by 
\begin{equation}
\varepsilon_{\pm}(q)=\pm [E_{\pm}(q)-E_{1/3}]-h,
\end{equation}
where $E_{+}(q)$ and $E_{-}(q)$ are the lowest total energy
at the sector $q$ for $S^z=S^{z}_{1/3}+1$ and $S^z=S^{z}_{1/3}-1$, respectively, with $h=0$. 
In Fig. \ref{fig:edgebb}(b) we show $\varepsilon_{\pm}$ for a closed system with $N_c=10$ and 12, and $h=(h_++h_-)/2$
for $\lambda=0.4$ (gapped magnon in the 1/3 -- plateau phase) and $\lambda=-0.5$ (gapless spinon in the LL phase). 
The expected \cite{Tsvelik90,Affleck91,Sorensen1993,Sachdev1994,PhysRevB.55.58} long-wavelength behavior is
also sketched with full lines. 
For $h_{-} < h < h_{+}$ (inside the 1/3 -- plateau), the excitations should obey a quadratic dispersion 
relation \cite{Tsvelik90,Affleck91,Sorensen1993,Sachdev1994,PhysRevB.55.58} 
\begin{equation}
\varepsilon_{\pm}(q)\rightarrow h_{\pm}\pm\frac{v_{\pm}^2}{2h_{\pm}}q^2-h\text{ as }q\rightarrow 0, 
\end{equation}
where $v_{\pm}$ are the spin-wave velocities (see discussion in Sec. \ref{sec:pd}).
For $\lambda=0.4$, shown in Fig. \ref{fig:edgebb}(b),  
a fitting (full lines) gives $v^2/2h \approx 0.61$ (0.62) for the  upper (lower) band.    
On the other hand, in the gapless LL phase, the upper and lower bands are joined at $q=0$, and 
the excitations follow a linear dispersion relation 
\begin{equation}
\varepsilon_{\pm}(q)\rightarrow \pm v_s|q|\text{ as }q\rightarrow0, 
\end{equation}
where $v_s$ is the spinon velocity.
\section{Boundary and magnon-magnon scattering lengths}
\label{sec:sca}
\subsection{Boundary scattering length for magnon excitations from the 1/3 -- plateau magnetization}
\label{secsec:bsca}
\begin{figure}
\includegraphics*[width=0.47\textwidth]{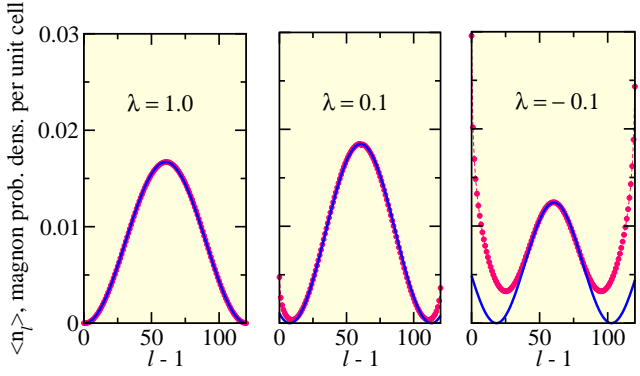}
\caption{Average magnon density $\langle n_l \rangle$ per unit cell along the chain for 
the extended one magnon excitation in the 1/3--plateau state. ({\color{red}$\bullet$}) DMRG results for $N_c=121$. The full line is a fitting of the DMRG data to the continuum limit expression for the probability density (far from the boundaries) of a particle in a box with a finite potential at the boundaries: $A\sin^2[\pi(x-a_b)/(N_c-2a_b)]$, 
where $a_b$ parameterizes the interaction with the boundaries, $A$ is a fitting parameter, and 
$x=l-1$. The fitting is done using the data in the range $x=45\ldots75$.  
}
\label{fig:magnondens}
\end{figure}

Here, we consider the average density profile of the isolated extended magnon excitation, obtained
from the magnetization change $\cn{3}\rightarrow\cn{4}$, as described in Sec. \ref{sec:edge}.
In our open chain, the bulk  magnon
lives on a squeezed chain with size \cite{Sca2} $N_c-2a_b$, where the
\textit{boundary scattering length} $a_b$
accounts for the repulsive ($a_b>0$) boundary potentials. Thereby far enough from the boundaries,
the bulk single-particle wavefunctions in the open chain can be written as \cite{Sca2} 
\begin{equation}
\psi_p(x)= \sqrt{A}\sin\left[\frac{p\pi (x-a_b)}{(N_c-2a_b)}\right],
\label{eq:singlep}
\end{equation}
where $p=1,~2,\ldots$ and $A$ is a constant. In Fig. \ref{fig:magnondens} we fit the DMRG data for the chain
with $N_c=121$ unit cells to the expression in Eq. (\ref{eq:singlep}) with $p=1$, 
and obtain $a_b=0.6$, $8.0$, and $18.0$, for $\lambda=1.0$, 0.1, and $-0.1$, respectively.

\subsection{Fully polarized plateau: insulator with trivial topology, and magnon-magnon scattering length}
\label{secsec:fpsca}
The fully polarized plateau is an example of a topological trivial insulator, 
with a Chern number $C_{3/2}=0$ (see discussion in Sec. \ref{sec:pd}). Thus, in an open chain, the fully polarized state does not have edge states.
Below we present the bulk magnon excitations from the fully polarized plateau, including the linear correction 
for the square-root law, and discuss the magnon density profile for two magnons in an open chain.
\begin{figure}
\includegraphics*[width=0.47\textwidth]{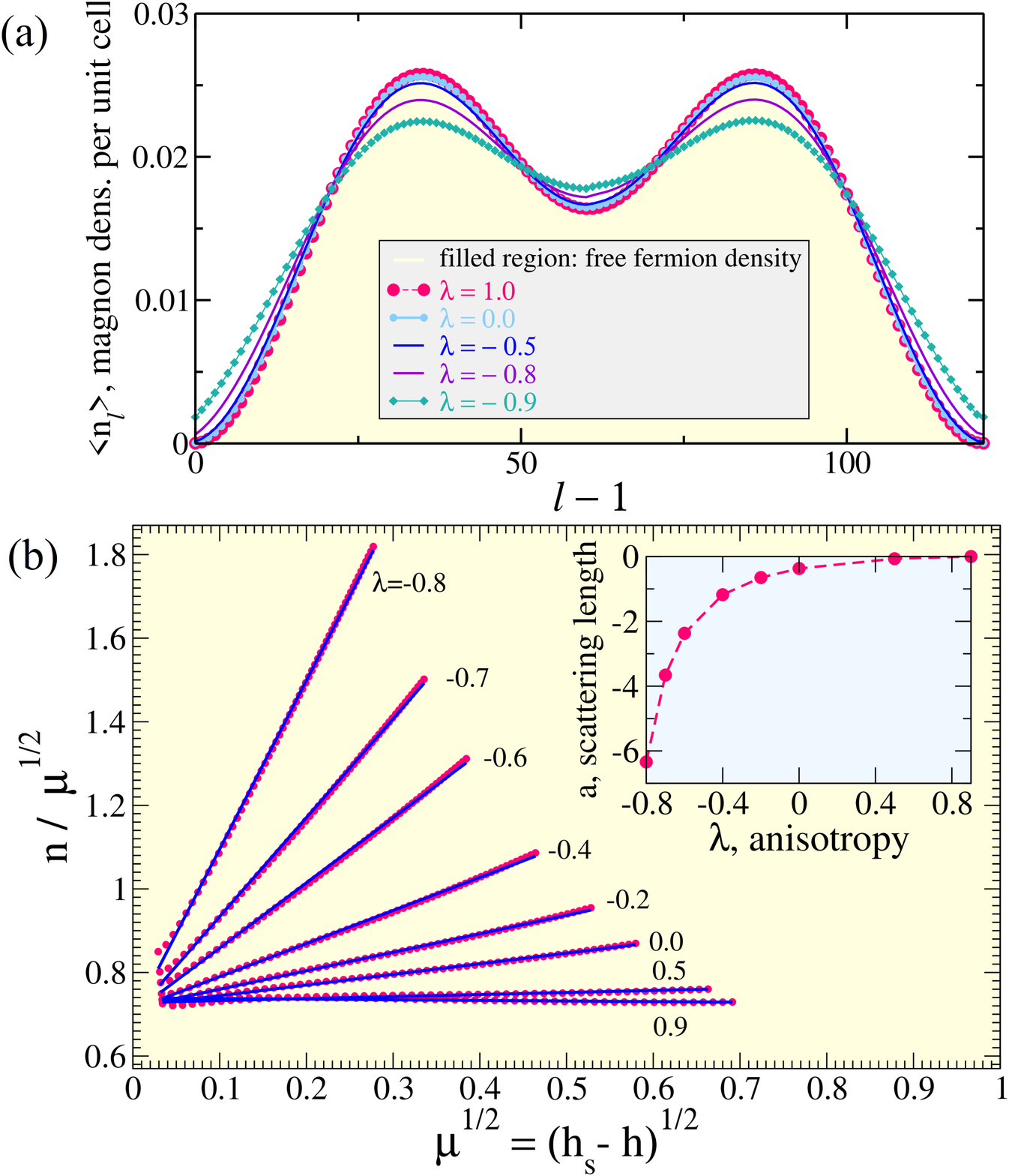}
\caption{Dilute magnon regime and the scattering length $a$, DMRG results for $N_c=121$. 
(a) Magnon density $\langle n_l\rangle$ along the chain for two magnons added to the FP state for the indicated values of $\lambda$.
(b) Average magnon density per unit cell $n$ for the FP -- plateau: $n=m_{FP}-m$, with $m_{FP}=(3/2)+(1/2N_c)$, 
as a function of $\mu^{1/2}$, where $\mu=h_s-h$ is the effective chemical potential and $h_s$ is the saturation field. Inset: scattering length $a$ derived from a fitting of the DMRG results to the expression of the effective fermion model with a linear correction: $n/\mu^{1/2}=\beta -\frac{4}{3}a\beta^2\mu^{1/2}$, with $\beta$ and the scattering length $a$ as fitting parameters.} 
\label{fig:bulk}
\end{figure}

In Fig. \ref{fig:bulk}(a) we present the two-particle average magnon density along the chain for the fully
polarized plateau, $\langle n_l\rangle$. For comparison, we show the free fermion density for two fermions in a
chain of size $N_c-1$ and vanishing boundary condition:
\begin{equation}
 \frac{2}{N_c-1}\left[\sin^2\left(\frac{\pi x}{N_c-1}\right)+\sin^2\left(\frac{\pi x}{N_c-1}\right)\right],
\label{eq:twofermion}
 \end{equation}
with $x=l-1$. We notice the absence of edge states in this case for $-0.9\leq\lambda\leq1.0$. A tiny departure 
from the free fermion result is observed as $\lambda\rightarrow -1$, the critical ferromagnetic point. The average magnon density increases at the boundaries with a decrease in the central region as $\lambda\rightarrow -1$. We explain it  
by noticing that if a $\Delta S^z = -1$ magnon is at a boundary $A$ site, with the other sites fully polarized, the  value of the longitudinal term of the energy is $-\lambda$. If the magnon is not at a boundary site, this energy term is $-4\lambda$ (at a $B_1$ or $B_2$ site) or $-2\lambda$ (at an $A$ site). Hence, for $\lambda<0$ the effect of the boundaries is represented by an attractive potential at the chain ends. However, while in Fig. \ref{fig:magnondens} we can observe a crossover between the profiles at the center and at the boundaries of the 
chain, this crossover is not evidenced in the density profiles shown in Fig. \ref{fig:bulk}(a). 

In the high-dilute limit of magnons near the $h_s(\lambda)$ line, the bulk magnon density per unit cell is given by
\begin{equation}
n=\sqrt{\frac{2 h_s \mu}{\pi^2 v^2}}, 
\end{equation}
with $n=m_{FP}-m$, $\mu=h_s-h$, and $v$ in Eq. (\ref{eq:vfp}).
Including the linear first correction \cite{Sca1,Sca2,affleck2004,Affleck2005,Sca3} to the square-root law, 
the magnon density becomes
\begin{equation}
 n=\sqrt{\frac{2 h_s}{\pi^2 v^2}}\sqrt{\mu}-a\frac{4}{3}\frac{2 h_s}{\pi^2 v^2}\mu,
 \label{eq:magnondens}
\end{equation}
where $a$ is the magnon-magnon scattering length, which can be positive or negative. For an infinite hard-core potential, $a>0$ and is equal to the core size, while $a<0$ for a repulsive delta-function potential. Hence, 
from the effective low-energy theory, we expect $a<0$.

In Fig. \ref{fig:bulk}(b), we show DMRG data for $n$ normalized by $\mu^{1/2}$ as a function of $\mu^{1/2}$ for 
$N_c=121$. The magnetization values shown range from $m=m_{FP}-(3/N_c)$ (three magnons) to $m=1$
(one magnon per unit cell). In order to obtain $a$ as a function of $\lambda$, 
we compare the DMRG data with the expression in Eq. (\ref{eq:magnondens}). 
In fact, from Eq. (\ref{eq:magnondens}), we find
\begin{equation}
 \frac{n}{\mu^{1/2}}=\beta-a\frac{4}{3}\beta^2 \mu^{1/2},
 \label{eq:fita}
\end{equation}
with 
\begin{equation}
\beta(\lambda)=\sqrt{\frac{2h_s}{\pi^2 v^2}}.
\label{eq:beta}
\end{equation}

We fit the full set of DMRG data in Fig. \ref{fig:bulk}(b) to Eq. (\ref{eq:fita}), for each $\lambda$ value, 
by considering $\beta$ and $a$ as fitting parameters.  Indeed, the relative departure between the values
of $\beta$ from the fitting and the ones obtained from Eq. (\ref{eq:beta}) ranges from 5\% to 10 \%. 
In Fig. \ref{fig:bulk}(b), we observe that $n/\mu^{1/2}$ is almost constant for $\lambda=0.9$, implying
the prevalence of the square-root behavior for these magnetization values. The scattering length $a$, shown
in the Inset, is $\approx 0$ for $\lambda=0.9$, and the hard-core boson or free fermion model is thus the best
effective theory. Notice that the value of $a$ decreases smoothly as $\lambda$ decreases 
and takes only negative values as expected for a $\delta$-function potential. 

\section{Summary and conclusions}
\label{sec:summary}

In summary, we use the density matrix renormalization group to discuss the phase diagram of 
the anisotropic AB$_2$ chain with an applied magnetic field. In particular, we reveal the locus of 
the magnon edge states, observed in finite size systems, inside the gap of the topological 1/3 -- plateau state. 
Besides, we use the transverse spin correlation functions to 
estimate the critical point of the Kosterlitz-Thouless transition: $\lambda_{KT}=-0.419\pm0.004$ and $h_{KT}=0.290\pm0.002$, such that we reach a better precision than known results. We also display the magnon distribution in 
the edge states and in the first extended state above the gap. Further, we follow the penetration of the edge states 
in the bulk as the 1/3 -- plateau gap closes. The gap closing is also 
accompanied by an effective squeezing of the chain, parameterized by a boundary scattering length. Considering the bulk states, we also use exact diagonalization to show the topological change in the dispersion relation of the excitations 
in the vicinity of the Kosterlitz-Thouless transition point. Furthermore, we studied the topologically trivial 
fully polarized plateau state. Since this insulating state is trivial, we show that the boundary magnon distributions in this case are distinct from that of the excitations from the topological 1/3 -- plateau state.
Particularly, we estimate the magnon-magnon scattering length as a function of the anisotropy 
and confirm that it provides a good correction (linear) to the square-root singularity in the
dilute regime of magnons. 
 
We expect that the reported features of the quantum many-body edge and extended states, and the rich 
phase diagram of the anisotropic Heisenberg AB$_2$ chain in a magnetic field, notably
the KT transition and the topological change of the excitations, will stimulate theoretical and experimental 
investigations in quasi-one-dimensional compounds exhibiting topological 1/3 magnetization plateaus, including 
ultra-cold optical lattice analogs. 

\begin{acknowledgments}
We acknowledge support from Coordena\c{c}\~ao de Aperfei\c{c}oamento de Pessoal de N\'{\i}vel Superior (CAPES),
Conselho Nacional de Desenvolvimento Cient\'{\i}fico e Tecnol\'ogico (CNPq), and Funda\c{c}\~ao de Amparo \`a Ci\^encia e
Tecnologia do Estado de Pernambuco (FACEPE), Brazilian agencies, including the PRONEX Program which is funded by
CNPq and FACEPE, APQ-0602-1.05/14. 
\end{acknowledgments}

\appendix
\begin{figure}
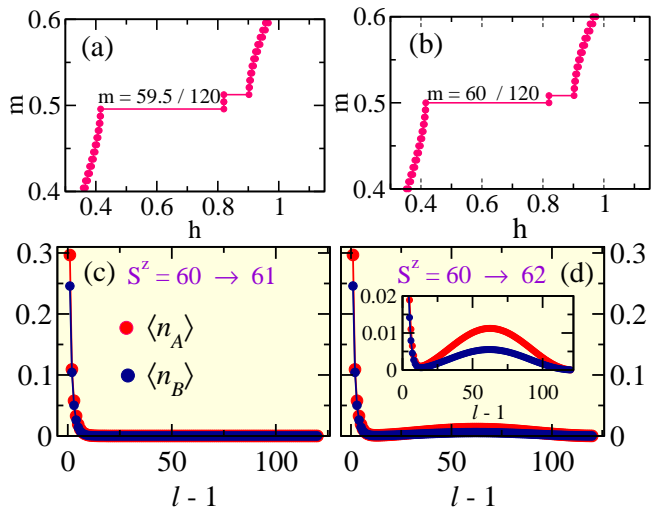

\includegraphics*[width=0.235\textwidth]{fig9a.eps}
\includegraphics*[width=0.235\textwidth]{fig9b.eps}
\includegraphics*[width=0.47\textwidth]{fig9cd.eps}
\caption{(a) and (b): DMRG results for $m(h)$ at $\lambda = 0.4$ 
in the vicinity of the 1/3 -- plateau magnetization for 
an even number of unit cells: $N_c=120$. In (a) we display results for one $A$ site at 
each boundary; 
while in (b) we consider one $A$ site at the left boundary and $B_1,B_2$ sites at the right boundary.
(c) One and (d) two 
magnon excitations above the magnetization $m=60/120$: average distribution at sites $A$, $\langle n_A\rangle$, and $B$,
$\langle n_B\rangle\equiv\langle n_{B_1}\rangle+\langle n_{B_2}\rangle$, as a function of the cell
position $l-1$ for the same boundary condition as in (b). 
}
\label{fig:distbc}
\end{figure}

\section{Magnetization for an even number of unit cells}
\label{sec:appendixA}
In Fig. \ref{fig:distbc}(a) we show the magnetization $m(h)$ for an even number of unit cells, $N_c=120$, 
$\lambda=0.4$, 
and the same boundary conditions used along the manuscript: one $A$ site at each boundary, with 
the 1/3 -- plateau magnetization at $m=(1/2)-(1/2N_c)=59.5/120$.  
This curve should be compared with that in Fig. \ref{fig:edgea}(a) for $N_c=121$. 
The physical features are essentially
identical to that of an odd number $N_c$, except that for even $N_c$ the minimum value of the spin 
is $S^z=1/2$, since the chain has an odd number of sites. 

In Fig. \ref{fig:distbc}(b) we show $m(h)$ for a chain with one $A$ site at the left boundary, and  
$B_1,B_2$ sites at the right boundary, for $N_c=120$. In this case, the system presents only one edge state in the left boundary, as shown in Fig. \ref{fig:distbc}(c) through the magnon distribution along the chain. We also 
show in Fig. \ref{fig:distbc}(d) the first extended magnon excitation. These figures should be compared with the excitations $\cn{1}\rightarrow\cn{2}$ and $\cn{1}\rightarrow\cn{4}$ in Fig. \ref{fig:edgea}(b). This behavior can be understood by noticing that there is a ``local'' distinction between $A$ and $B_1,B_2$ sites. This is equivalent to the presence of distinct local potentials for $A$ and $B_1,B_2$ sites, such that this difference inhibits the
edge state in the $B_1,B_2$ boundary.  

\section{Average local magnetizations}
\label{sec:appendixB}

\begin{figure}[htb]
\includegraphics*[width=0.47\textwidth]{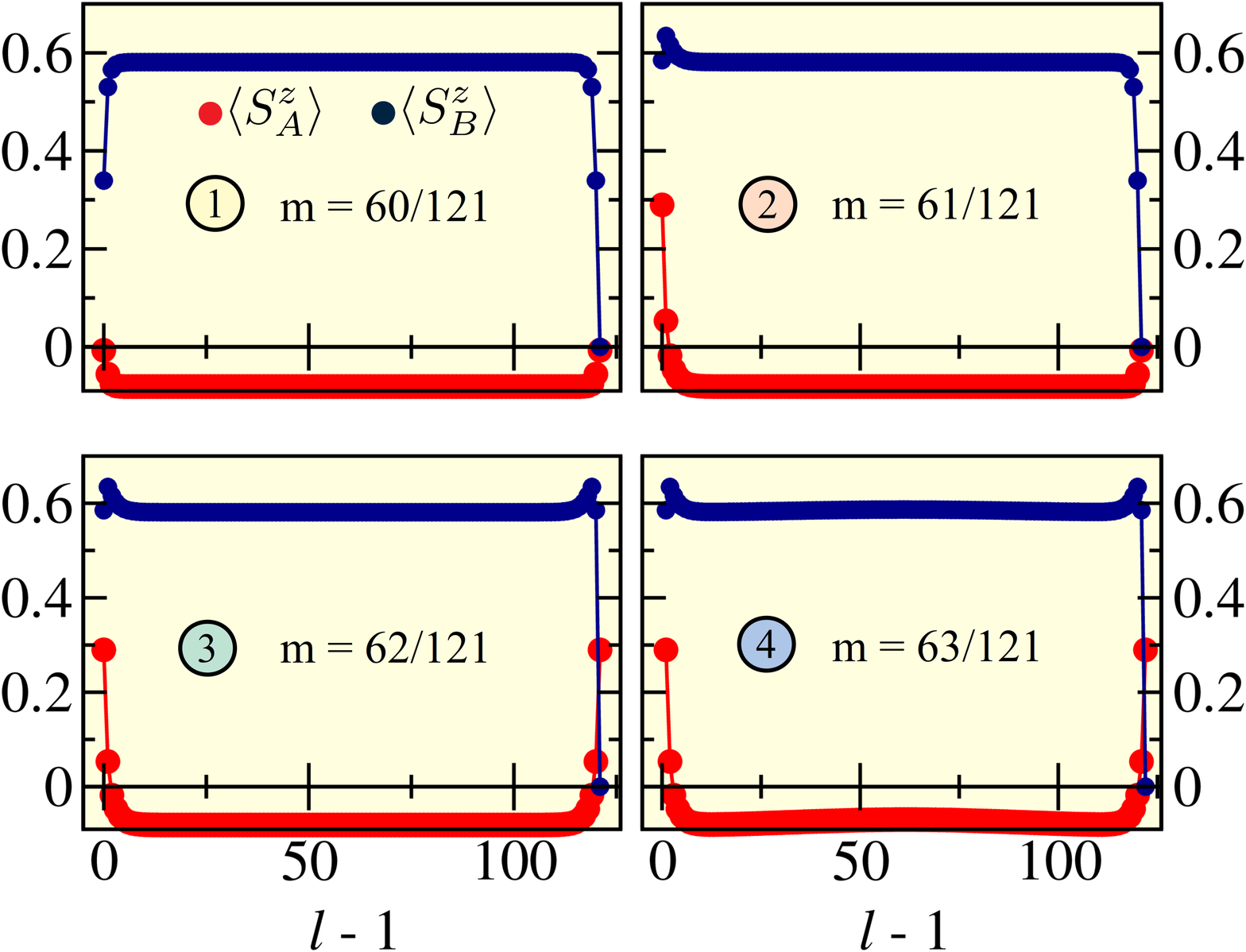}
\caption{(a) and (b): DMRG results for the average spin distribution at sites $A$, $\langle S^z_A\rangle$, and $B$,
$\langle S^z_B\rangle\equiv\langle S^z_{B_1}\rangle+\langle S^z_{B_2}\rangle$, as a function of cell
position $l-1$ at the indicated $m$-states: \cn{1} ($m=60/121$), \cn{2} ($m=61/121$), \cn{3} ($m=62/121$), and 
\cn{4} ($m=63/121$) for a chain with $N_c=121$ and $\lambda=0.4$.
}
\label{fig:lmag}
\end{figure}

In Fig. \ref{fig:lmag} we present the average magnetizations at sites $A$, $\langle S^z_A\rangle$, and at $B$ sites, 
$\langle S^z_B\rangle=\langle S^z_{B_1}\rangle+\langle S^z_{B_2}\rangle$  as a function of cell position 
for $\lambda=0.4$. These magnetizations were used to build the curves for the average magnon distributions 
shown in Fig. \ref{fig:edgea}(b), in which case the edge magnon states are highlighted. 
This is one of the two degenerate states, which is chosen by the renormalization procedure, as 
explained in the main text.

\bibliography{RefPRB}

\end{document}